\documentclass{article}
 \usepackage{graphicx}
 \def\be{\begin{equation}}
 \def\ee{\end{equation}}
 \newcommand{\jhep}[3]{J.~High Energy Phys.\ #1 (#2) #3}
 \newcommand{\hepph}[1]{\texttt{hep-ph/#1}}
 \newcommand{\heplat}[1]{\texttt{hep-lat/#1}}
 \newcommand{\prd}[3]{Phys.~Rev.~D#1 (#2) #3}
 \newcommand{\plb}[3]{Phys.~Lett.~B#1 (#2) #3}
 \newcommand{\lsim}{\,\raise-0.3ex\hbox{$\sim$}\kern-.7em\hbox{$^>$}\,}
 \newcommand{\ssim}{\,\raise-0.3ex\hbox{$\sim$}\kern-.7em\hbox{$^<$}\,}
\begin{document}

\title{Comment on `Pressure of hot QCD at large $N_{f}$'}

\author{A.~Peshier\\
Institut f\"{u}r Theoretische Physik, Universit\"{a}t Giessen\\
35392 Giessen, Germany}
\maketitle

\vskip 5mm

\begin{abstract} \noindent
It is argued why quasiparticle models can be useful to describe the
thermodynamics of hot QCD excluding, however, the case of a large
number of flavors, for which exact results have been calculated by Moore.
\end{abstract}

\noindent
In a recent paper~\cite{Moore02}, Moore considered the thermodynamics
of hot QCD with a large number of quark flavors, $N_{f} \gg N_c$.
Although this case is physically not directly relevant, it is still
interesting to study the limit $N_{f} \rightarrow \infty$ as the
next-to leading order result in the $1/N_{f}$ expansion of the
pressure can be calculated exactly as a function of the effective
coupling $G^2 = N_{f} g^2/2$ in this model.\footnote{Note that the
resulting theory is not asymptotically free and leads to a Landau
pole, so it is not defined in a strict sense. However, as argued
in~\cite{Moore02}, the model can be understood as an effective theory
with a controllable ambiguity which is small as long as the Landau
pole is much larger than any other scale in the problem.}  Therefore,
it is suggestive to use the large-$N_{f}$ limit of QCD as a testing
ground for the convergence of perturbative results~\cite{ZhaiK} or
resummation improved methods~\cite{AnderBS,BlaizIR3,Peshi01}. From the
strong-coupling behavior of the pressure at large $N_{f}$ it was also
concluded in~\cite{Moore02} that the quasiparticle model~\cite{PKPS96}
cannot be a good description for `real' QCD near the QCD
transition. In the following it is argued why such a conclusion has to
be taken with some care.

To start with, a sketchy derivation of the large-$N_{f}$ pressure is
given within the $\Phi$-derivable approach. Setting the volume of the
system to one, the QCD pressure for arbitrary $N_{f}$ can be expressed
in terms of the full quark and gluon propagators~\cite{BlaizIR3,
Peshi01} (the ghost contribution and the subtraction of the vacuum
pressure are not shown explicitly),
\be
 p
 =
 {\rm Tr}\left[ \ln\left(-S^{-1}\right)+\Sigma S \right]
 -\frac12 {\rm Tr}\left[ \ln\left(-D^{-1}\right)+\Pi D \right]
 +\Phi[S,D] \, .
 \label{eq: Omega}
\ee
The traces are taken over the 4-momentum and, accordingly, over the
flavor, spin, color and Lorentz indices. $\Phi$ is the sum of all
2-particle irreducible bubble diagrams. The self-energies are obtained
by cutting a corresponding line in $\Phi$. In the large-$N_{f}$
limit, $\Phi$ reduces to the 1-gluon exchange diagram for each flavor
and contributes $\frac12$Tr$\Pi D$ to $p$, which compensates the
second term in the gluon contribution. While the gluon self-energy is
of the order $G^2$, the quark self-energy is of the order
$G^2/N_{f}$. Thus, the leading term in the $1/N_{f}$ expansion of
$p$ is given by the free quark pressure, $p_{\,lo} = {\rm
Tr}\ln(-S_0^{-1}) \sim N_{f}$. The term of the order $N_{f}^0$
comes from the remaining gluon contribution~\cite{Moore02},
\be
  p_{\rm nlo} = -\frac12 {\rm Tr} \ln\left(-D^{-1}\right).
\ee
Moore's numerical computation for massless quarks shows that $p_{\rm
nlo}$ first decreases with increasing $G^2$ as expected, but then
starts to increase again, exceeding even the free value $p_{\rm
nlo}(G^2=0)$ before the coupling becomes eventually so large that the
position of the Landau pole reaches the order of the temperature and
the model becomes meaningless.\footnote{After submitting this article,
Moore's numerical results for $p_{\rm nlo}$ as published
in~\cite{Moore02} were corrected in~\cite{ippmr}. While the pressure
still has a minimum, it now exceeds the free pressure only at a value
of the coupling where it is already sensitive to the Landau pole
ambiguity.} Clearly, such a non-monotonic behavior cannot be described
by quasiparticles whose masses rise with the coupling.

\begin{figure}[hbt]
\begin{minipage}[t]{57mm}
  \centerline{\includegraphics[scale=0.55]{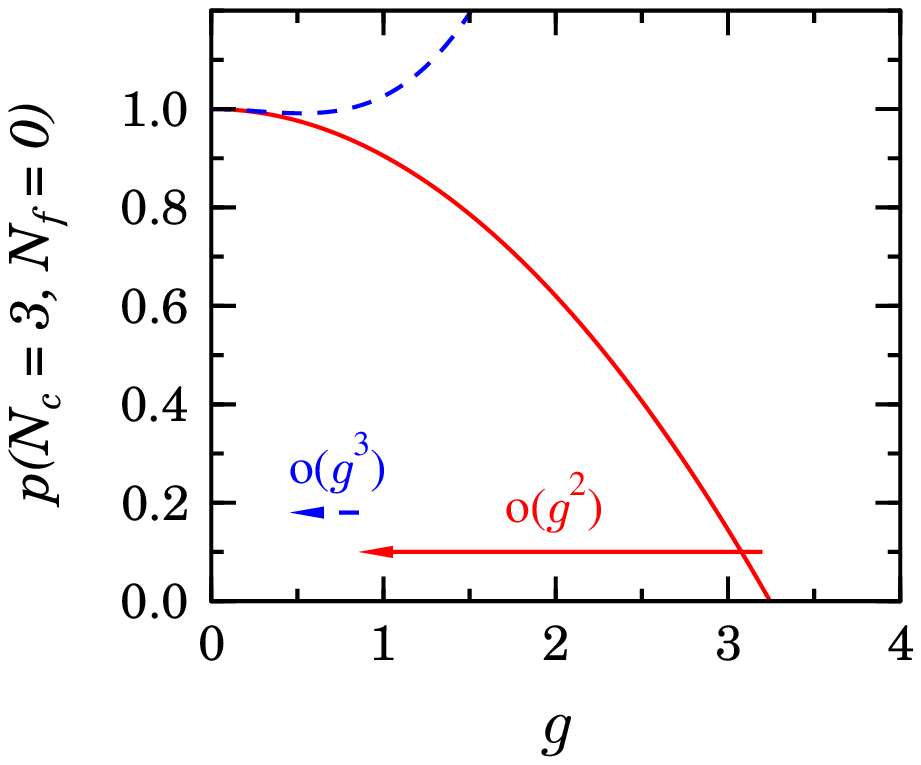}}
  \vskip-3mm
  \caption{The perturbative approximations, $p^{(2)}$ and $p^{(3)}$,
    of the pressure for $N_{\!f}=0$, in units of the free value.
    \label{fig: Nf=0}}
\end{minipage}
\hfill
\begin{minipage}[t]{57mm}
  \centerline{\includegraphics[scale=0.55]{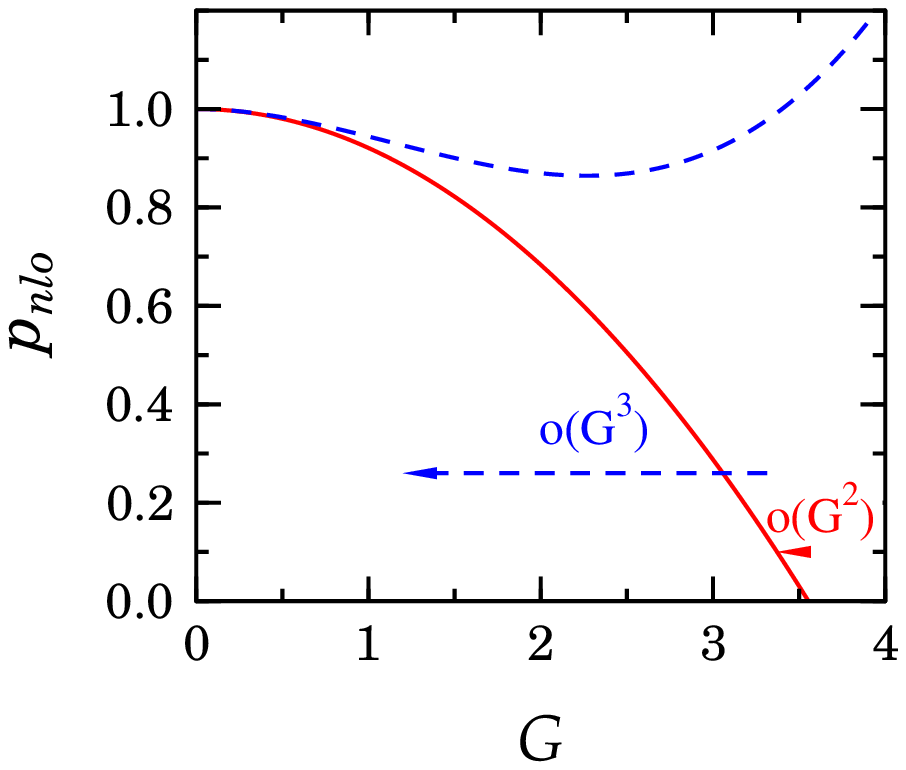}}
  \vskip-3mm
  \caption{The approximations $p_{nlo}^{(2)}$ and $p_{nlo}^{(3)}$ of the
    large-$N_{\!f}$ pressure $p_{nlo}$, scaled by the free
    value.
    \label{fig: large Nf}}
\end{minipage}
\end{figure}

To understand the surprising behavior of the large-$N_{f}$ pressure,
and whether it might be relevant for real QCD, it is instructive to go
back to an arbitrary number of flavors. Then, in dimensionless units
and up to terms of order $g^4$, the perturbative expansion of the
pressure reads~\cite{ZhaiK}
\be
 p^{(3)}(N_c,N_{f})
 =
 \frac74\, \frac{d_f}{d_a}
 + 1
 - 5 \left( N_c + \frac54 N_{f} \right) \bar{g}^2
 + \frac{80}{\sqrt3}\,
    \left( N_c + \frac12\, N_{f}\right)^{\!3/2}\! \bar{g}^3 \,,
 \label{eq: p3}
\ee
with $d_f = N_c N_{f}$, $d_a = N_c^2-1$, and $\bar{g} = g/(4\pi)$.
The normalization is such that the free gluon contribution is one.
Note that to this order the result does not depend explicitly on the
renormalization scale. Aside from the fact that in general the terms
beyond ${\cal O}(g^5)$ in the expansion of $p$ cannot be calculated by
perturbative methods, the series is not convergent. The best one can
hope for is some sort of an asymptotic expansion. Such a series,
however, should be truncated at a certain order, depending on the
expansion parameter, to give the best approximation possible.
Typically, one should truncate the series at the $n$th term when its
modulus becomes as large as that of the preceding contribution, see,
e.\,g.,~\cite{GradR}.
Applying this prescription to the expansion~(\ref{eq: p3}) leads to
the estimate that $p^{(3)}$ is a reasonable approximation for $\bar g$
smaller than
\be
  \bar g_3
  =
  \frac{\sqrt3}{16}\,
  \frac{N_c+\frac54\, N_{f}}{(N_c+\frac12\, N_{f})^{3/2}} \, .
\ee
$p^{(3)}$ has a minimum at $\bar{g}^\star = \frac23\, \bar g_3$ (which
therefore might be physically relevant) with a relative depth of
\be
  \delta
  =
  1 - \frac{p^{(3)}(\bar{g}^\star)}{p^{3}(0)}
  =
  \frac5{144(4+7d_f/d_a)}
  \left( \frac{N_c+\frac54\, N_{f}}{N_c+\frac12\, N_{f}} \right)^3.
\ee
For $\bar g > \bar g_3$, the ${\cal O}(g^2)$ approximation, $p^{(2)}$,
is expected to be more adequate until it obviously also becomes
meaningless at
\be
  \bar g_2
  =
  \left( \frac15\, \frac{4+7d_f/d_a}{4N_c+5N_{f}} \right)^{1/2} .
\ee
The values of $g_2$, $g_3$ and $\delta$ for $N_c =3$ and several
numbers of $N_{f}$ are given in table~\ref{table1}, and the representative case
$N_{f} = 0$ is illustrated in figure~\ref{fig: Nf=0}.

\begin{table}[hbt]
\begin{center}
\begin{tabular}{c||c|c|c||c}
         & $N_{\!f}=0$ & $N_{\!f}=3$ & $N_{\!f}=6$ & large $N_{\!f}$
\\ \hline
$g_2$    &  3.2        & 3.7         & 3.9         & $G_2 = 3.5$
\\
$g_3$    &  0.79       & 0.96        & 0.97        & $G_3 = 3.4$
\\
$\delta$ & $0.0087$    & $0.0099$    & $0.0094$    & $0.14$
\end{tabular}
\end{center}
\vskip-3mm\caption{
  The values of the coupling indicating the validity range of
  the quadratic approximation $p^{(2)}$ of the pressure for several
  $N_{f}$ and of $p_{\rm nlo}$ in the large-$N_{f}$ limit, respectively,
  and the depth $\delta$ of the minimum for the cubic approximation.}
\end{table}
Due to the estimated large validity range of the approximation
$p^{(2)}$ one can expect that the monotonically decreasing behavior at
large coupling is qualitatively correct for the exact pressure.  This
is indeed what is observed in QCD lattice calculations with a small
number of flavors~\cite{KarscLP}, where the pressure decreases
monotonically with the temperature and reaches a rather small value at
the transition temperature $T_c$.  Although it is not relevant for the
large coupling which is of primary interest here, it is mentioned that
the minimum of the approximation $p^{(3)}$ is very shallow, and that
the ${\cal O}(g^4)$ contribution to $p$ is for reasonable choices of
the renormalization scale always smaller than the ${\cal O}(g^3)$
contribution.\footnote{In figure~\ref{fig: Nf=0}, the lower bound of
the validity interval of $p^{(3)}$ is estimated from the magnitude of
the 4th and 5th order contribution to $p$.}  The latter observation
might indicate that the perturbative series is not an asymptotic
expansion in the strict sense.

Considering now the large-$N_{f}$ limit, one can read off the
perturbative expansion of $p_{\rm nlo}$ from eq.~(\ref{eq: p3}) after
subtracting the term $\sim N_{f}$; with $\bar G = G/(4\pi)$,
\be
  p_{\rm nlo}^{(3)}
  =
  1 - \frac{25}2\, \bar G^2 + \frac{80}{\sqrt3}\, \bar G^3 \, .
  \label{eq: p3_nlo}
\ee
The perturbative structure of $p_{\rm nlo}$ is rather different from
the case discussed above. Although the ${\cal O}(G^2)$ approximation
vanishes at a similar value of the (effective) coupling, the
estimated (as before) validity range of $p_{\rm nlo}^{(2)}$ is tiny, see
table~\ref{table1} and figure~\ref{fig: large Nf}.
Therefore, one cannot expect $p_{\rm nlo}$ to become small at large
coupling --- contrary to the pressure at small $N_{f}$. Again, this
expectation is in line with the exact result~\cite{Moore02}. Numerically, the
expression $p_{\rm nlo}^{(3)}$ turns out to be a reasonable
approximation\footnote{It is noted that the ${\cal O}(G^3)$
contribution is (for relevant values of the coupling) always larger
than the ${\cal O}(G^4)$ contribution which is larger than the
contribution of the order $G^5$.}  for values of $G$ as large as the
coupling at the minimum of $p_{\rm nlo}$, while $p_{\rm nlo}^{(2)}$
never seems to be the optimal approximation.

The truncation prescription for asymptotic series implies that the
terms of a higher than a certain order, which depends on the size of
the expansion parameter, almost cancel each other. For the pressure of
QCD with a few flavors, in the region of physical interest, near the
QCD transition, this occurs already after the leading order correction
in $g^2$. Hence, a quasiparticle model like~\cite{PKPS96}, which
incorporates the leading perturbative correction and resums higher
order terms only partly to ensure thermodynamic consistency, appears
to be justified.  The situation is different for the large-$N_{f}$
limit of QCD where the pressure $p_{\rm nlo}$ has a peculiar
perturbative structure. This fact is physically plausible: Here the
cubic plasmon term, i.\,e., the correction to the leading exchange
contribution due to screening, is small since quarks screen less than
gluons. Consequently, this term is relevant at much larger coupling
than in the physical case.

In conclusion, Moore's argument --- the quasiparticle
model~\cite{PKPS96} cannot describe the minimum of $p_{\rm nlo}$ at
large $N_{f}$ and is thus not expected to be a good description of the
physical case either --- would only be convincing if such a pronounced
minimum also occurred in real QCD; otherwise the large-$N_{f}$ limit
is governed by different physics. By the argument that in both cases
the strong-coupling behavior of the pressure is reflected in the
appropriately interpreted perturbative results, or directly by
contrasting the numerical results (lattice data for real QCD), this,
however, seems indeed to be the case.
\\[3mm] \noindent
{\bf Acknowledgments:} I thank S.~Leupold for comments on the manuscript.
This work is supported by BMBF.

\end{document}